\documentclass[twocolumn,pra,showpacs,superscriptaddress]{revtex4-1}
\usepackage{amssymb}
\usepackage{mathrsfs}
\usepackage[pdftex]{graphicx}
\usepackage[caption=false]{subfig}
\usepackage{mathtools}
\usepackage[pdftex,colorlinks=true]{hyperref}
\DeclareGraphicsExtensions{.pdf,.jpg,.png}
\usepackage[toc,page,title,titletoc,header]{appendix}
\usepackage{tikz}
\usetikzlibrary{decorations.pathmorphing}

\begin{document}

\title{Single-photon scattering with counter rotating wave
  interaction}

\author{Qi-Kai He}

\affiliation{Institute of Physics, Beijing National Laboratory for
  Condensed Matter Physics, Chinese Academy of Sciences, Beijing
  100190, China}

\affiliation{School of Physical Sciences, University of Chinese
  Academy of Sciences, Beijing 100190, China}

\author{Wei Zhu}

\affiliation{Institute of Physics, Beijing National Laboratory for
  Condensed Matter Physics, Chinese Academy of Sciences, Beijing
  100190, China}

\author{Z.~H. Wang}

\affiliation{Center for Quantum Sciences and School of Physics,
  Northeast Normal University, Changchun 130024, China}

\author{D.~L. Zhou}

\email{zhoudl72@iphy.ac.cn}

\affiliation{Institute of Physics, Beijing National Laboratory for
  Condensed Matter Physics, Chinese Academy of Sciences, Beijing
  100190, China}

\affiliation{School of Physical Sciences, University of Chinese
  Academy of Sciences, Beijing 100190, China}

\date{\today}

\begin{abstract}
  Recent experiments have pushed the studies on atom-photon
  interactions to the ultrastrong regime,  which motivates the
  exploration of physics beyond the rotation wave approximation.
  Here we study the single-photon scattering on a system composed by
  a coupling cavity array with a two-level atom in the center
  cavity, which, by varying two outside coupling parameters,
  corresponds to a model from a supercavity QED to a waveguide QED
  with counter-rotating wave (CRW) interaction. By applying a
  time-independent scattering theory based on the bound states in
  the scattering region, we find that the CRW interaction obviously
  changes the transmission valley even in the weak atom-cavity
  coupling regime; In particular, the CRW interaction leads to an
  inelastic scattering process and a Fano-type resonance, which is
  directly observed in the crossover from the supercavity QED case
  to the waveguide QED case. Predictably, our findings provide the
  potential of manipulating the CRW effects in realistic systems.
\end{abstract}


\maketitle

\section{Introduction}

Recent experiments on diverse systems, such as circuit
QEDs~\cite{NDH+2010,FLM+2010,BHH+2016}, 2d electron
gases~\cite{Scalari1323}, spiropyran molecules~\cite{SHGE2011}, and
semiconductor quantum wells~\cite{GCS+2012}, have pushed the research
on photon-atom interactions to the ultrastrong regime, where the
coupling is so strong that the rotating wave approximation
(RWA)~\cite{1443594} is not valid any longer, and the effects from the
counter rotating wave interaction (CRW) can not be neglected.

In the RWA, the interacting photon and atom only exchange their
excitations, thus the total excitation is conserved, which greatly
simplifies the underlying physics and the theoretical treatments. The
CRW interaction makes the total excitation not conserved, which makes
the relative phenomena and the calculations become complex. To solve
the calculation problem, several theoretical methods are introduced,
such as, the generalized rotating-wave approximation
(GRWA)~\cite{PhysRevLett.99.173601}, the analytical solution in the
Bargmann space~\cite{PhysRevLett.107.100401}, and the numerical method
based on matrix product states
(MPS)~\cite{doi:10.1080/14789940801912366,1367-2630-8-12-305}.

To study the effects from the CRW interaction, it is convenient to
investigate the single-photon scattering with an (artificial) atom in
a one dimensional supercavity (SC) or
waveguide~\cite{shen_coherent_2005}. In the RWA, a one dimensional
waveguide model is composed of coupling cavity array (CCA) with one
cavity locating a two-level atom is firstly proposed in
Ref.~\cite{PhysRevLett.101.100501}. An extension to present a SC-QED
model is given in Ref.~\cite{PhysRevA.90.043828}, where the concept of
SC is borrowed from Ref.~\cite{PhysRevA.78.053806}. A natural problem
is to extend the above models to the ultrastrong coupling regime. In
fact, the CCA waveguide model including the CRW interaction is firstly
studied in Ref.~\cite{PhysRevA.86.023824} by using the GRWA\@. Then a
remarkable result from this model in
Ref.~\cite{PhysRevLett.113.263604} is to discover an inelastic
scattering process by using wave packet scattering simulation based on
MPS algorithm.

In this article, we extend the SC-QED model to the ultrastrong regime,
and apply the time-independent scattering theory to study the single
photon transmission spectrum. Further more, our method is a unified
frame to study the crossover from the SC-QED model to the waveguide
QED model. In particular, we will show how the CRW interaction affect
the single photon transmission in our model. For example, we find that
there is an obvious effect in the photon transmission even in the weak
coupling regime; the inelastic scattering also occurs in the SC-QED
model as that predicted in the waveguide QED
model~\cite{PhysRevLett.113.263604}.

The rest of this paper is built up as follows. In Sec.~\ref{sec:2}, we
introduce our model and numerically calculate the bound states of the
SC, which is further confirmed by the Brillouin-Wigner perturbation
theory~\cite{BWPT2010}. Based on the single-photon scattering mechanism
given in Sec.~\ref{sec:2B}, the numerical results of the single-photon
scattering process in our model are presented in
Sec.~\ref{sec:3}, showing how the CRW interaction affects the single
photon transmission. In Sec.~\ref{sec:4}, we give some discussions and
draw the conclusions.

\section{Model And Basic Scattering Processes\label{sec:2}}

\subsection{The model and Hamiltonian\label{sec:2A}}

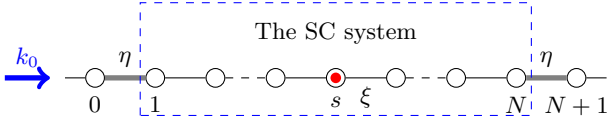
\begin{figure}[!htbp]
  \begin{tikzpicture}[node-s1/.style={circle,draw,inner sep=2.5pt}]
    \node[node-s1,label=below:$0$] (cl1) at (-0.8,0) {};
    \node[node-s1,label=below:$1$] (c1)
    at (0,0) {};
    \node[node-s1] (c2) at (0.8,0) {};
    \node[node-s1] (c3) at
    (1.6,0) {};
    \node[node-s1,label=below:$s$]
    (c4) at (2.4,0) {};
    \node[node-s1] (c5) at (3.2,0) {};
    \node[node-s1] (c6) at (4.0,0) {};
    \node[node-s1,label=below:$N$] (c7) at (4.8,0) {};
    \node[node-s1,label=below:$N+1$] (cr1) at (5.6,0) {};
    \draw (c1)--(c2) (c3)--(c4)--node[below]{$\xi$}(c5) (c6)--(c7) (-1.2,0)--(cl1)
    (cr1)--(6.0,0);
    \draw[dashed] (c2)--(c3) (c5)--(c6);
    \draw[gray,line width=2pt] (cl1)--node[above,black]{$\eta$}(c1)
    (c7)--node[above,black]{$\eta$}(cr1);
    \fill[red] (2.4,0) circle (2pt);
    \draw[dashed,blue] (-0.2,-0.5) rectangle (5.0,1.0);
    \node at (2.4,0.6) {The SC system};
    \draw[->,blue,line width=2pt]
    (-2.0,0)--node[above]{$k_{0}$}(-1.4,0);
  \end{tikzpicture}
  \caption{(Color online). Demonstration of our model. A single
    photon (blue right arrow) with the momentum $\vec{k}_{0}$ injects
    from the left end of the 1D infinite CCA and the transmission
    spectrum is measured on the right side.}
  \label{fig:FullModel}
\end{figure}

As shown in Fig.~\ref{fig:FullModel}, the system we consider contains
a one-dimensional coupled cavity array (CCA) with infinite length and
a two-level atom, where each cavity is represented by an empty circle
and the two-level atom is represented by a red solid circle. The
cavities in the CCA are labeled by integers in increasing order from
left to right. The photonic hopping strengths between the neighbouring
cavities $l$ and $l+1$ are $\eta$ for $l=0$ or $l=N$ and $\xi$ for
others. When $\eta\ll\xi$, the CCA between $l=1$ and $l=N$ forms a
multi-mode cavity, which will be named as a SC. The two-level atom
that locates in the $s$-th cavity of the SC (Let $N$ be odd, $s \equiv \frac{N+1}{2}$), together with the SC, constructs a cavity-QED system,
denoted as the SC system in Fig.~\ref{fig:FullModel}.

The Hamiltonian of our system is written as (we set $\hbar=1$)
\begin{equation}
  \hat{H} = \hat{H}_{S}+ \hat{H}_{L} + \hat{H}_{R} + \hat{H}_{LS} +
  \hat{H}_{SR},
  \label{eq:1}
\end{equation}
where
\begin{subequations}
  \begin{eqnarray}
    \hat{H}_{S} & = & \omega_{c} \sum^{N}_{j=1} \hat{a}^{\dagger}_{j}
                      \hat{a}_{j} - \xi \sum^{N}_{j=2} (
                      \hat{a}^{\dagger}_{j} \hat{a}_{j-1} + h.c. ) +
                      \frac{\omega_{a}}{2} \hat{\sigma}_{z} \nonumber\\
                  & & \quad {} + g \hat{\sigma}_{x}(
                      \hat{a}^{\dagger}_{s} +
                      \hat{a}_{s}), \label{eq:2a}\\
    \hat{H}_{L}& =& \sum^{0}_{j=-\infty} [ \omega_{c}
                    \hat{a}^{\dagger}_{j} \hat{a}_{j} - \xi
                    (\hat{a}^{\dagger}_{j} \hat{a}_{j-1} +
                    \hat{a}^{\dagger}_{j-1} \hat{a}_{j} ) ],\\
    \hat{H}_{R} &=& \sum^{\infty}_{j=N+1}
                    [ \omega_{c} \hat{a}^{\dagger}_{j} \hat{a}_{j} - \xi
                    (\hat{a}^{\dagger}_{j} \hat{a}_{j+1} +
                    \hat{a}^{\dagger}_{j+1} \hat{a}_{j} ) ], \\
    \hat{H}_{LS} &=& -\eta (\hat{a}^{\dagger}_{0} \hat{a}_{1} +
                     \hat{a}^{\dagger}_{1} \hat{a}_{0}),\\
    \hat{H}_{SR} &=& -\eta (\hat{a}^{\dagger}_{N} \hat{a}_{N+1} +
                     \hat{a}^{\dagger}_{N+1} \hat{a}_{N}).
  \end{eqnarray}
  \label{eq:2}
\end{subequations}
Here $\hat{H}_{S}$ is the Hamiltonian of the SC system, $\hat{H}_{L}$
($\hat{H}_{R}$) describes the left (right) channel which is used to
input (output) photons, and $\hat{H}_{LS}$ ($\hat{H}_{SR}$) describes
the interaction between the SC system with the left (right) channel.
The operator $\hat{a}^{\dagger}_{j}$ ($\hat{a}_{j}$) is the photon
creation (annihilation) operator for the $j$-th cavity,
$\hat{\sigma}_{-}=|g\rangle\langle e|$
($\hat{\sigma}_{+}={\hat{\sigma}_{-}}^{\dagger}$) is the atomic
lowering (raising) operator. $\omega_{c}$ is the mode frequency of
cavities, and $\omega_{a}$ is the energy level splitting of the atom.

Our model can be regarded as a direct generalization of the model in
Ref.~\cite{PhysRevA.90.043828}, where the RWA approximation is made.
In addition, when $\eta=\xi$, our model becomes the one studied in
Refs.~\cite{PhysRevLett.113.263604} and \cite{PhysRevA.86.023824}.

The basic task in this paper is to investigate scattering behavior for
the incident single-photon from the left channel. To this end, we
should firstly analyze the intrinsic energy-level structure of the
scatterer (i.e., the SC system which is shown in the blue dashed frame
in Fig.~\ref{fig:FullModel}), since the bound states of the SC
system will modify the elastic scattering and induce the inelastic
scattering for the incident photon.

\subsection{Bound states and basic scattering processes\label{sec:2B}}

The interaction term between
the two-level atom and the SC in the Hamiltonian of the SC system
$\hat{H}_{S}$ in Eq.~\eqref{eq:2a} is
\begin{align}
  \hat{H}_{\rm{int}} & = g \hat{\sigma}_{x}( \hat{a}^{\dagger}_{s} +
                       \hat{a}_{s} ) \nonumber\\
                     & = \hat{H}^{\rm{RW}}_{s}+\hat{H}^{\rm{CRW}}_{\rm{int}},
\end{align}
where
\begin{align}
  \label{eq:4}
  \hat{H}^{\rm{RW}}_{\rm{int}} & = g (\hat{\sigma}_{+} \hat{a}_{s} +
                                 \hat{a}^{\dagger}_{s}
                                 \hat{\sigma}_{-}), \\
  \hat{H}^{\rm{CRW}}_{\rm{int}} & = g (\hat{\sigma}_{+}
                                  \hat{a}^{\dagger}_{s} +
                                  \hat{a}_{s} \hat{\sigma}_{-}).\label{eq:5}
\end{align}
Here $\hat{H}^{\rm{RW}}_{\rm{int}}$ is the `rotating wave' term and
$\hat{H}^{\rm{CRW}}_{\rm{int}}$ is the CRW term. As we know, the
effect of $H_{\rm{int}}^{\rm{CRW}}$ can be safely neglected whenever
$g\ll\{\omega_a,\omega_c\}$, that is, the rotating wave approximation
is applicable in this case.
In the rotating wave approximation, the excitation number
$\hat{N}_{\rm{ext}}=\sum_{j=1}^{N}\hat{a}_j^{\dagger}\hat{a}_j+(\hat{\sigma}_z+1)/2$
is conserved. When the CRW term can not be neglected, the excitation
number is not conserved any longer, however the parity operator
$\hat{P}=(-1)^{\hat{N}_{\rm{ext}}}$ satisfies $[\hat{P},\hat{H}_S]=0$,
which is a $\text{Z}_2$ symmetry.

Then we use the numerical exact diagonalization
algorithm~\cite{zhang_exact_2010} to diagonalize the Hamiltonian
$\hat{H}_{S}$, and rewrite it as
\begin{equation}
  \label{eq:6}
  \hat{H}_{S} = \sum_{m} \epsilon_{m} |\varphi_{m}\rangle
  \langle\varphi_{m}|,
\end{equation}
where $m\in\{1,2,\cdots\}$,
$\epsilon_{1} \leq \epsilon_{2} \leq \cdots$, $\epsilon_{m}$ and
$| \varphi_{m} \rangle$ are the $m$-th eigenenergy and the
corresponding eigenvector, respectively.

For most eigenstates of $\hat{H}_S$, photons will be distributed in
the whole SC, that is, to form extended states. However, there also
exists some bound states due to the interaction with the two-level
atom in the $s$-th cavity. As will be explained later, these bound
states play an essential role in the inelastic scattering in our
problem. Thus it is worthy to explore the origin of these bound
states.

\begin{figure}[!htbp]
  \centering
  \includegraphics[width=0.45\textwidth]{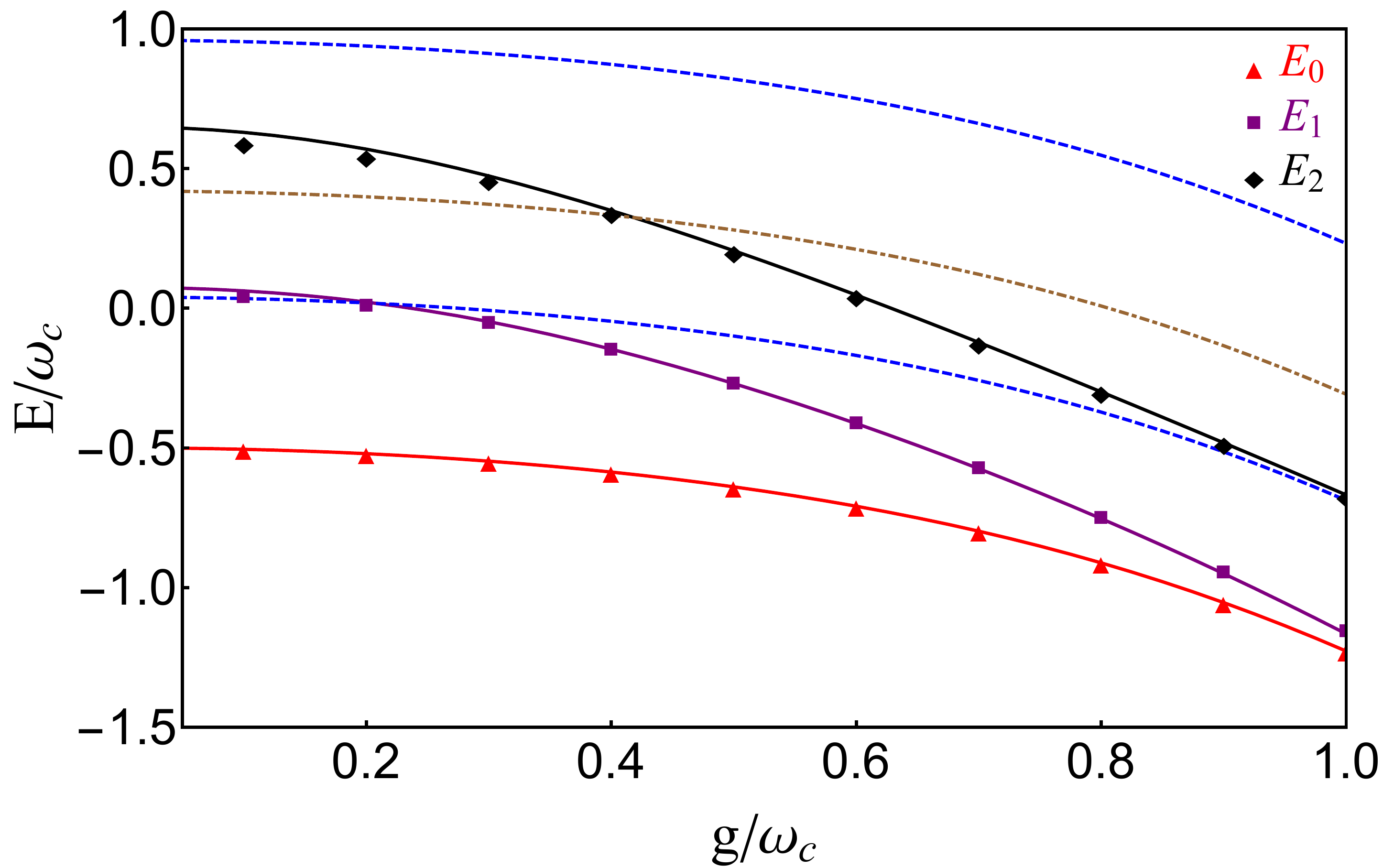}
  \caption{(Color online). Energies of bound states for different
    coupling strength $g$. The points represent bound state energies
    obtained by the Brillouin-Wigner perturbation theory (BWPT) when
    the length of the SC is long enough. The solid lines represent
    results obtained via numerical diagonalization for the SC with
    $N = 7$. The total energy $E_{\rm{in}}$ of the single-photon
    scattering process is in the range between two blue dashed lines.
    The brown dash-dotted line stands for the upper limit of $E_{2}$
    for the inelastic scattering process. Here we take
    $\omega_{a} / \omega_{c} = 1$ and $\xi / \omega_{c} = 0.23$ in
    units of $\omega_{c}=1$.}
  \label{fig:BSenergy}
\end{figure}

As interpreted in Appendix~\ref{app:A1}, we use the Brillouin-Wigner
perturbation theory (BWPT)~\cite{BWPT2010} to obtain the bound states
$|\psi_{m}\rangle$ and the corresponding energies $E_{m}$. Obviously,
$\{ E_{m} \} \varsubsetneq \{ \epsilon_{n} \}$. Then we may use these
results to select the bound states from all the eigenstates of
$\hat{H}_{S}$ by the direct numerical diagonalization. The numerical
results on the three lowest eigenenergies of bound states as a
function of coupling strength $g$ are shown in
Fig.~\ref{fig:BSenergy}.

Compared to the numerical diagonalization method, the utilization of
the BWPT can save the time to identify the bound states and the
internal storage in calculation under the same condition. Thus we use
the BWPT to get the three lowest energy levels of the bound states,
which are shown by points in Fig.~\ref{fig:BSenergy}, when the length
of the SC is long enough. The convergence of our results is examined
in Appendix~\ref{app:A2}. As shown in Fig.~\ref{fig:BSenergy}, we find
that the ground state energies with $g$ obtained from numerical
diagonalization of the Hamiltonian~\eqref{eq:2a} for the SC with $N=7$
agree well with those corresponding points. The energy of
$| \psi_{2} \rangle$ is in close agreement with that obtained via the
BWPT below its upper limit for the inelastic scattering process. In
our numerical calculations, the excitation $\hat{N}_{\rm{ext}}$ is cut
off at a given number $N_{\rm{ext}}=7$, which is found to be
sufficient to guarantee the convergence of numerical results. We also
find that all three bound states have certain parity and obey
$\hat{P}|\psi_0\rangle = |\psi_0\rangle , \hat{P}|\psi_1\rangle =
-|\psi_1\rangle , \hat{P}|\psi_2\rangle = |\psi_2\rangle$. In other
words, $|\psi_0\rangle$ and $|\psi_2\rangle$ are the states with even
parity and $|\psi_1\rangle$ is that with odd parity.

With the extended and bound states of the SC system, we can
furthermore study the single-photon scattering in the full system. The
single-photon scattering process in our model can be formulated
as follows. Initially, we inject a photon with momentum $k_{0}$ and
prepare the SC system at its ground state $|\psi_{0}\rangle$ (the
lowest bound state). Absorbing the input photon with frequency
$\omega_{\rm{in}}=\omega_c-2\xi\cos k_{0}$, the scatterer (the SC
system) will jump to the extended eigenstates, which are not stable
due to its coupling with the left and right channels. Then, a photon
will be emitted with the carrying frequency $\omega_{\rm{out}}$, and
the scatterer will pass to some bound state $|\psi_m\rangle$.

According to the energy conservation in the scattering process, we
have
\begin{equation}
  E_{\rm{in}} = E_{0} + \omega_{\rm{in}} = E_{m} + \omega_{\rm{out}}.
  \label{conservation}
\end{equation}
It implies that $\omega_{\rm{out}}=\omega_{\rm{in}}$ for the elastic
scattering process ($E_m=E_0$) while
$\omega_{\rm{out}}\neq\omega_{\rm{in}}$ for the inelastic scattering
process ($E_m\neq E_0$). Under appropriate parameter conditions (as
shown below), the two scattering processes will occur simultaneously.
Furthermore, according to the conservation of the parity and the
energy, the single-photon inelastic scattering for $m=2$ occurs.
Therefore we get $\omega_{\rm{in}}=E_2-E_0+\omega_{\rm{out}}$ and the
condition for the single-photon inelastic scattering is
\begin{equation}
  \omega_{\rm{in}} \geq E_2 - E_0 + \omega_c - 2\xi.
  \label{eq:8}
\end{equation}

With Eq.~\eqref{eq:8}, we get the upper limit of $E_{2}$ for the
inelastic scattering process as $E_{0} + 4\xi$, which is shown by a
brown dash-dotted line in Fig.~\ref{fig:BSenergy}. This implies that
the length $N=7$ of the SC is enough for our investigation of
single-photon inelastic scattering.

\section{Numerical Results And Analysis\label{sec:3}}

Now let us study the single-photon scattering process based on our
model in the regimes of $\eta/\xi \in [0,1]$, in particular the two
cases $\eta / \xi\ll 1$ and $\eta / \xi\approx 1$. In the
single-photon scattering process, we consider only the photon states
in the left channel and the right channel up to one photon, which is a
good approximation when the length of the SC is sufficient long
such that the multi-photon process occur only in the cavities near the
atom~\cite{PhysRevA.86.023824}.
In this approximation, the time-independent scattering state can be
written as
\begin{align}
  |\Psi_{s}\rangle & =  |\Phi_{\rm{k_{0}}}\rangle + r_{\text{e}}
                     |\Phi^{*}_{\rm{k_{0}}}\rangle + t_{\text{e}} |\Theta_{\rm{k_{0}}}\rangle + \sum_{m}
                     d_{m} |\text{vac};\varphi_{m};\text{vac}\rangle \nonumber\\
                   & \quad {} + r_{\text{in}}
                     |\Phi^{*}_{\rm{k_{2}}}\rangle + t_{\text{in}}
                     |\Theta_{\rm{k_{2}}}\rangle
                     \label{eq:9}
\end{align}
with
\begin{align}
  |\Phi_{\rm{k_{i}}}\rangle & = \sum^{0}_{j=-\infty} e^{ijk_{i}}
                              |j;\psi_{i};\text{vac}\rangle, \\
  |\Theta_{\rm{k_{i}}}\rangle & = \sum^{\infty}_{j=N+1} e^{ijk_{i}}
                                |\text{vac};\psi_{i};j\rangle,~\label{eq:10}
\end{align}
which satisfies the Schr$\ddot{o}$dinger equation
\begin{equation}
  \hat{H} |\Psi_{s}\rangle = E_{\rm{in}} |\Psi_{s}\rangle,
  \label{eq:12}
\end{equation}
where the coefficients $r_{\text{e}}$ ($r_{\text{in}}$) and
$t_{\text{e}}$ ($t_{\text{in}}$) in Eq.~\eqref{eq:9} represent
respectively the reflection and transmission amplitudes in the elastic
(inelastic) scattering channel, $d_{m}$ is the probability amplitude
for the system in the state
$|\text{vac};\varphi_{m};\text{vac}\rangle$, and the eigenenergy
$E_{\rm{in}} = E_{0} + \omega_{c} - 2\xi\cos k_{0}$. Substituting
Eq.~\eqref{eq:9} into Eq.~\eqref{eq:12}, we get a set of equations
\begin{subequations}
  \begin{eqnarray}
    \xi r_{\text{e}} e^{-ik_{0}} - \eta \sum_{i=0} d_{i} \langle
    \psi_{0}|\hat{a}_{1}|\varphi_{i}\rangle & = & -\xi e^{ik_{0}},
                                                  \label{eq:13a} \\
    \xi t_{\text{e}} e^{iNk_{0}} - \eta \sum_{i=0} d_{i} \langle
    \psi_{0}|\hat{a}_{N}|\varphi_{i}\rangle & = & 0,
                                                  \label{eq:13b} \\
    \xi e^{-ik_{2}} r_{\text{in}} - \eta \sum_{i=0} d_{i}
    \langle \psi_{2}|\hat{a}_{1}|\varphi_{i}\rangle & = & 0,
                                                          \label{eq:13c} \\
    \xi e^{iNk_{2}} t_{\text{in}} - \eta \sum_{i=0} d_{i}
    \langle \psi_{2}|\hat{a}_{N}|\varphi_{i}\rangle & = & 0, \label{eq:13d}
  \end{eqnarray}
  \label{eq:13}
\end{subequations}
and
\begin{equation}
  \begin{split}
    & (E_{in} - \epsilon_{j}) d_{j} + \eta r_{\text{e}} \langle
    \varphi_{j} | \hat{a}^{\dagger}_{1} | \psi_{0} \rangle + \eta
    r_{\text{in}} \langle \varphi_{j} | \hat{a}^{\dagger}_{1} |
    \psi_{2} \rangle \\
    + & \eta t_{\text{e}} e^{i(N+1)k_{0}} \langle \varphi_{j} |
    \hat{a}^{\dagger}_{N} | \psi_{0} \rangle + \eta t_{\text{in}}
    e^{i(N+1)k_{2}} \langle \varphi_{j} | \hat{a}^{\dagger}_{N} |
    \psi_{2}\rangle \\
    = & -\eta \langle \varphi_{j} | \hat{a}^{\dagger}_{1} | \psi_{0}
    \rangle.
    \\
  \end{split}
  \label{eq:14}
\end{equation}

By numerically solving Eqs.~\eqref{eq:13} and \eqref{eq:14}, we will
obtain the reflection and transmission flow in the elastic scattering
channel as $J_{\rm{R,e}}=|r_{\text{e}}|^{2}$ and
$J_{\rm{T,e}}=|t_{\text{e}}|^{2}$, respectively. Due to the different
photon momentum in elastic and inelastic channels, the reflection and
transmission flow in the inelastic scattering channel should be
written as
$J_{\rm{R,in}}= \left| r_{\text{in}} \right|^{2} \sin k_{2}/\sin
k_{0}$ and
$J_{\rm{T,in}} = \big{|} t_{\text{in}} \big{|}^{2} \sin k_{2}/\sin
k_{0}$~\cite{PhysRevA.89.053813}. Then the flow conservation relation
is expressed as
$J_{\rm{R,e}}+J_{\rm{T,e}}+J_{\rm{R,in}}+J_{\rm{T,in}}=1$.

\subsection{Numerical results for $\eta \ll \xi$\label{sec:3A}}

In the regime of $\eta\ll\xi$, the SC system described by the
Hamiltonian $H_S$ couples to the left and right channels weakly.
Therefore, our whole system can be regarded as a microscopic
one-dimensional Cavity-QED model without the RWA, whose transmission
peaks of single-photon scattering correspond to the eigenstates
$\{| \varphi_{m} \rangle\}$ that satisfy
$\langle \varphi_{m} | \hat{a}^{\dagger}_{1} | \psi_{0} \rangle \neq
0$.

\begin{figure}[!htbp]
  \centering
  \subfloat[$g/\omega_{c} = 0.001$.]{
    \includegraphics[width=0.24\textwidth]{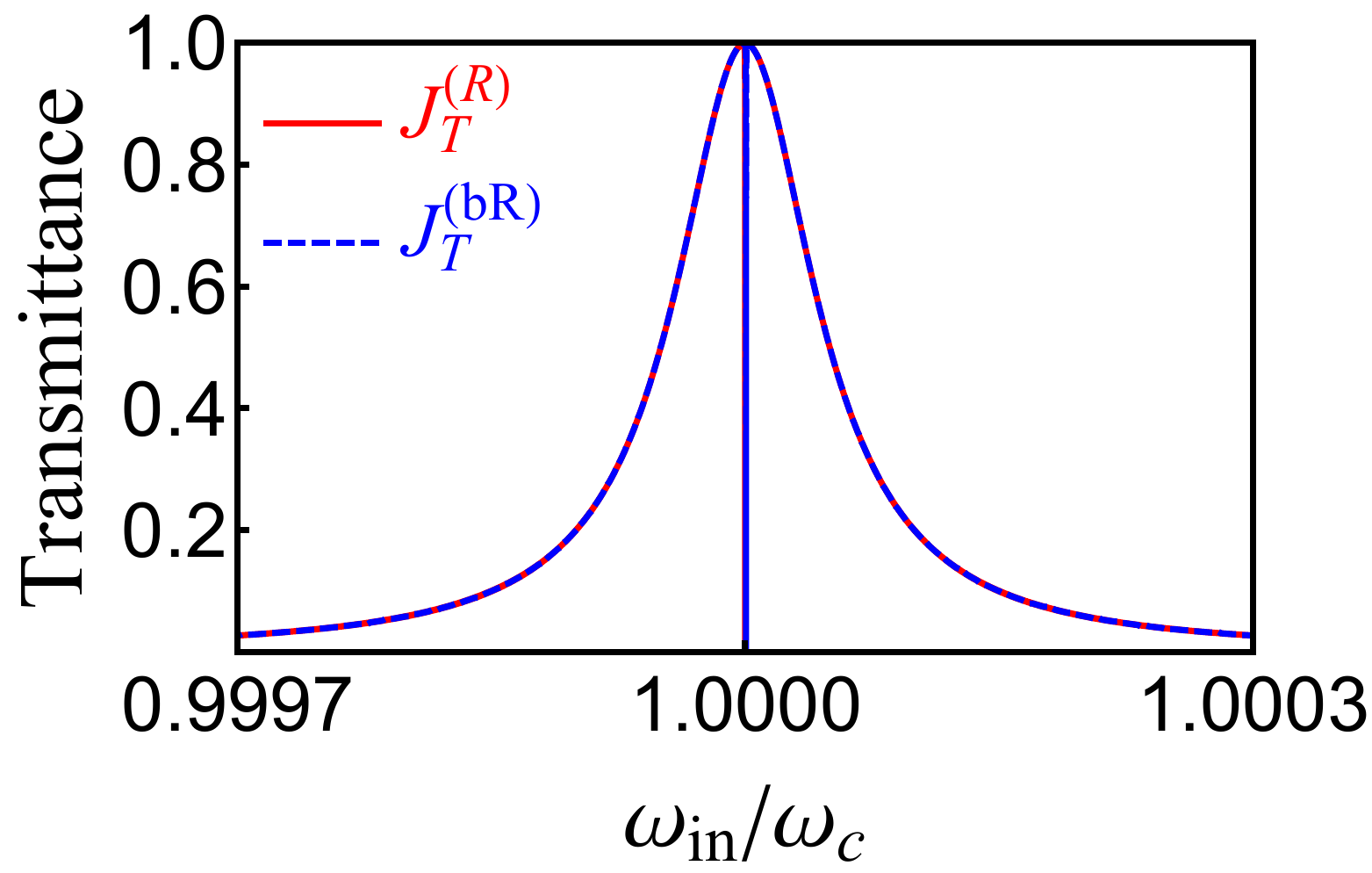}~\label{fig:Cg001} }
  \subfloat[Zoom in of \protect\subref{fig:Cg001}.]{
    \includegraphics[width=0.24\textwidth]{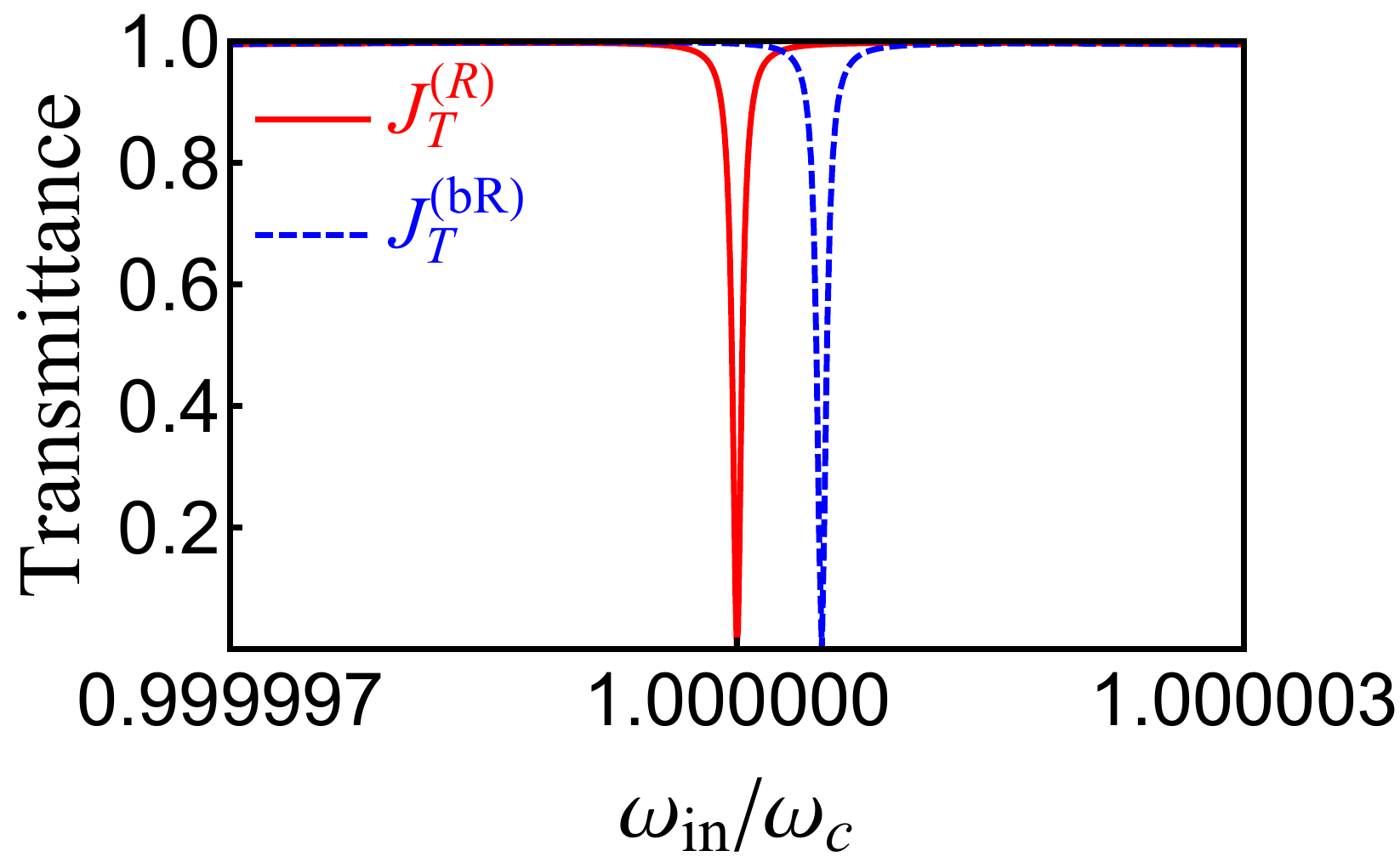}~\label{fig:Cg001Zoom}
  } \\
  \subfloat[$g/\omega_{c} = 0.01$.]{
    \includegraphics[width=0.24\textwidth]{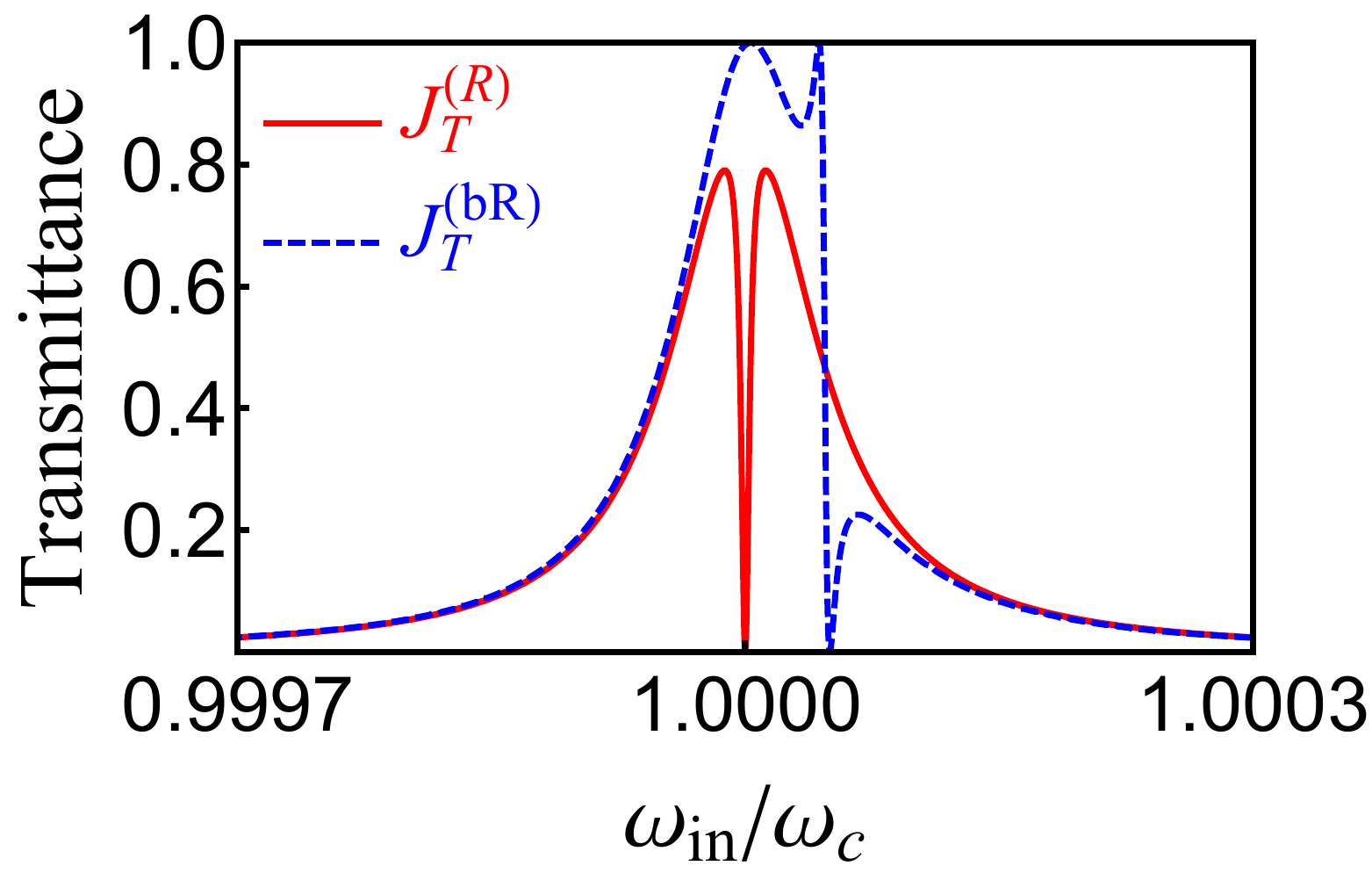}~\label{fig:Cg01Zoom} }
  \subfloat[$g/\omega_{c} = 0.05$.]{
    \includegraphics[width=0.24\textwidth]{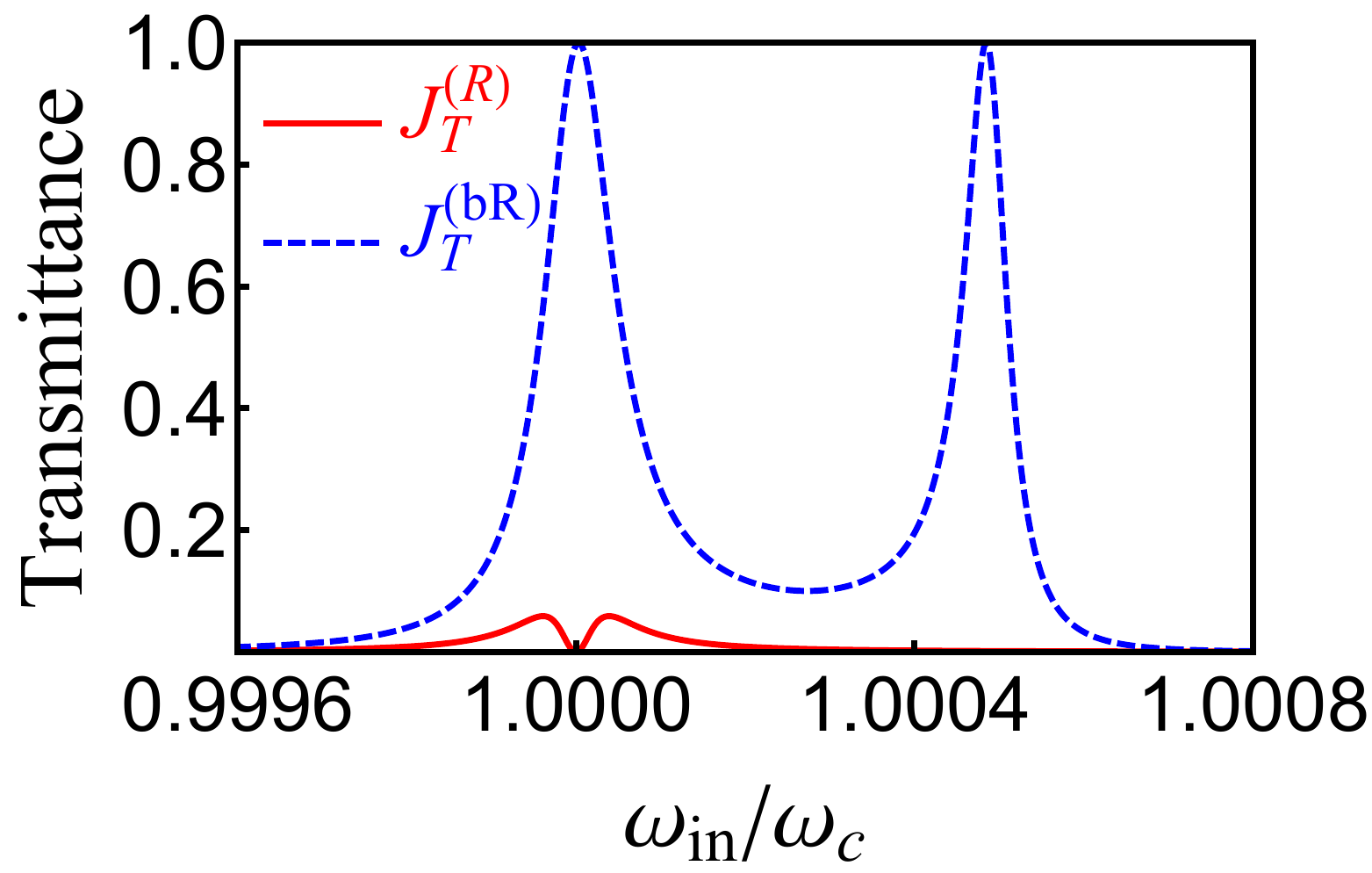}~\label{fig:Cg05Zoom} }
  \caption{(Color online). The transmittance as a function of
    $\omega_{\rm{in}}$ at $\omega_{\rm{in}} = \omega_{c}$. The red
    solid line represents the transmittance $J^{\rm{(R)}}_{\rm{T}}$ in
    the RWA while the blue dashed line stands for
    $J^{\rm{(bR)}}_{\rm{T}}$ beyond the RWA\@. Here we take the length
    of the scatterer $N=7$ and $\eta/\omega_{c} = 0.002$,
    $\xi/\omega_{c} = 0.04$.}~\label{fig:Cgsmallxi}
\end{figure}

When $\xi \ll \omega_{c}$, we have
$E_0+\omega_{\rm{in}}-E_2<\omega_{c}-2\xi$, which implies that only
the elastic scattering process occurs due to the energy conservation.
In this regime, a transmission valley induced by destructive
interference between two transmission channels in the RWA was found in
Ref.~\cite{PhysRevA.90.043828}. A natural problem is to investigate
whether and how the CRW term affects the transmission valley.

To this end, we compare the transmission rates in the cases with and
without the CRW term $\hat{H}_{\text{int}}^{\text{CRW}}$ when the
incident photon frequency $\omega_{\text{in}}$ is near $\omega_{c}$,
where the atom is located in the node of the resonant mode of the
SC\@. In Fig.~\ref{fig:Cgsmallxi}, we plot the elastic transmission
flow $J_{\rm{T,e}}$ as a function of the frequency of incident photon
for $g/\omega_{c}=0.001, 0.01, 0.05$ in \ref{fig:Cg001},
\ref{fig:Cg01Zoom}, \ref{fig:Cg05Zoom} respectively. In the RWA, the
transmission valley becomes lower and lower as $g$ increases, and the
position of the valley is invariant with $g$. The phenomena in the RWA
can be understood as follows. In the RWA, the eigenenergy of the
eigenstate formed by the atom coupling with the non-resonant modes of
the SC is invariant, which directly leads to the invariant position of
the transmission valley. As $g$ increases, this eigenstate has larger
decay into the outside channels, together with the destructive
interference mechanism, which explain why the transmission valley
becomes lower and lower.

When consider the CRW term, we find that when $g/\omega_{c}=0.001$,
there is a negligible effect of the CRW term in the transmission
valley. In fact, it has a small displace of the valley position as
shown in Fig.~\ref{fig:Cg001Zoom}. When $g/\omega_{c}=0.01$, the
transmission becomes asymmetric obviously. When $g/\omega_{c}=0.05$,
there appear two strong transmission peaks. The above phenomena
originate from the CRW term induced eigenenergy displacement of the
eigenstate formed by the atom coupling with the non-resonant modes,
which is possible to be sufficiently large with respect to the decay
rate of the resonant mode of the SC\@.

Moreover, we have also investigated the transmission spectrum when the
atom is located in the antinode of the empty SC, it exhibits the
normal Rabi splitting shape as expected, the RWA and full Hamiltonian
give almost the same results with the same parameters $g$ as above. It
means that the CRW term does not contribute significantly to the
dynamics of the system when the atom is not located in the node of the
resonant mode of the SC, in contrary to the case when the atom is
located in the node.

\begin{figure}[!htbp]
  \centering
  \subfloat[The total transmittance $J_{\text{T}}$.]{
    \includegraphics[width=0.45\textwidth]{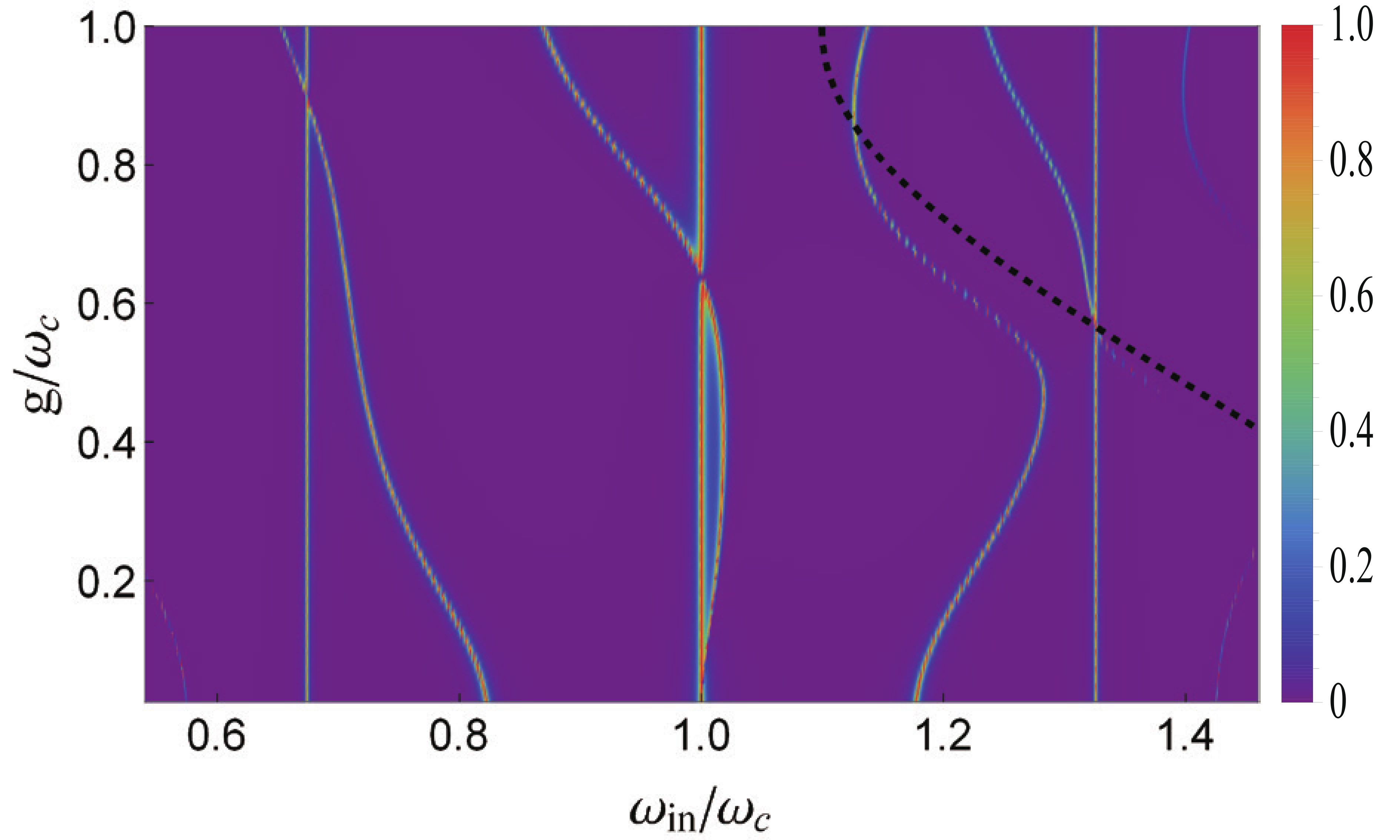}
    \label{fig:smalleta-T(a)}} \\
  \subfloat[The inelastic transmittance $J_{\text{T,in}}$.]{
    \includegraphics[width=0.45\textwidth]{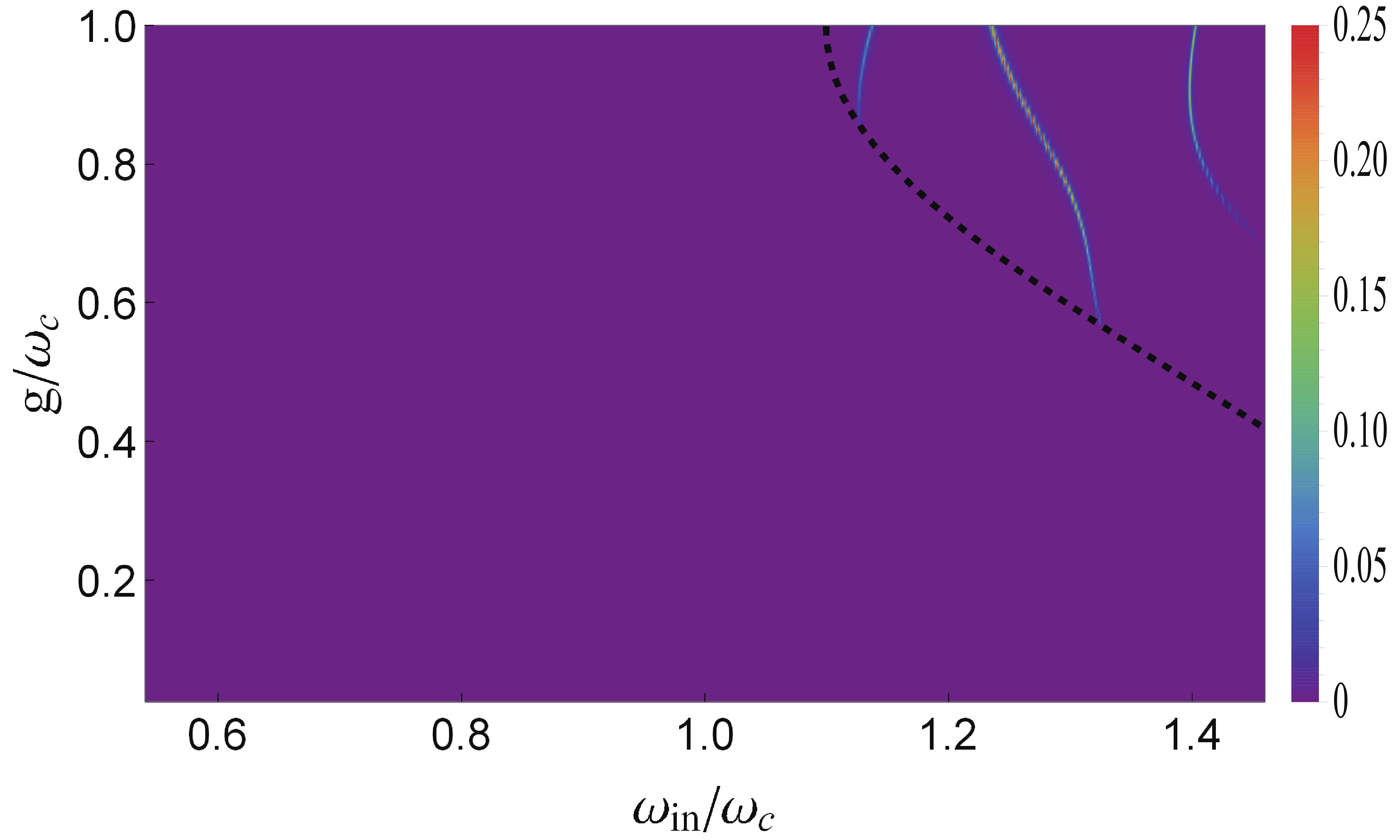}
    \label{fig:smalleta-Tin(b)}}
  \caption{(Color online). Transmittance as a function of incident
    photon frequency $\omega_{in}$ and $g$ for
    $\eta = 0.03\omega_{c},~\xi =0.23\omega_{c}$. The black dashed
    line marks the estimated frequency given by Eq.~\eqref{eq:8}.}
  \label{fig:TeTin-SC-eta-xi}
\end{figure}

Now, let us come to the parameter regime in which $\xi$ is in the same
order of $\omega_{c}$. In Fig.~\ref{fig:TeTin-SC-eta-xi} the total
transmission spectrum $J_{\rm{T}}=J_{\rm{T,e}}+J_{\rm{T,in}}$ and the
inelastic one $J_{\rm{T,in}}$ are shown for the SC-QED model. In
Fig.~\ref{fig:smalleta-Tin(b)}, we observe the inelastic scattering
when the condition Eq.~\eqref{eq:8} is satisfied  and
$J_{\rm{T,in}}$ exceeds $10\%$ for larger $g/\omega_{c}$ which
indicates that it can be easily observed in experiments.
Furthermore, these inelastic scattering peaks must be associated with
an extended state $| \varphi_{m} \rangle$ which satisfies
$\langle \psi_{2} | \hat{a}_{N} | \varphi_{m} \rangle \langle
\varphi_{m} | \hat{a}^{\dagger}_{1} | \psi_{0} \rangle \neq 0$.

In Fig.~\ref{fig:smalleta-T(a)}, we find that the
location of some peaks are the same which implies that the energy gaps
between the associated states and the ground states are not changed
for different $g$. To find the corresponding states, we rewrite the
Hamiltonian~\eqref{eq:2a} into
\begin{equation}
  \begin{split}
    \hat{H}_{S} =& \sum_{k} [ \omega_{c} - 2\xi \cos(\frac{k\pi}{N+1})
    ] \hat{b}^{\dagger}_{k} \hat{b}_{k} + \frac{\omega_{a}}{2}
    \hat{\sigma}_{z} \\
    & +\sum_{k} G_{k} \hat{\sigma}_{x} ( \hat{b}^{\dagger}_{k} +
    \hat{b}_{k} ),
  \end{split}
  \label{eq:15}
\end{equation}
where
\begin{align}
  \hat{b}^{\dagger}_{k} & = \sqrt{\frac{2}{N+1}} \sum^{N}_{j=1}
                          \sin(j \frac{k\pi}{N+1}) \hat{a}^{\dagger}_{j}, \label{eq:16} \\
  G_k & =g \sqrt{\frac{2}{N+1}} \sin(\frac{k\pi}{2}).
        \label{eq:17}
\end{align}

For $k'$ satisfying $G_{k'}=0$, we have
\begin{equation}
  \label{eq:18}
  \hat{H}_{S} b^{\dagger}_{k'} |\psi_{0}\rangle= \left(E_{0} + \omega_{c} -
    2\xi \cos\left(\frac{k' \pi}{N+1}\right)\right) b_{k'}^{\dagger} |\psi_{0}\rangle.
\end{equation}

In the case of $N=7$, $G_{k'}=0$ will give $k'=2,4,6$, and the energy
gaps of the states $b^{\dagger}_{k'} |\psi_{0}\rangle$ with respect to
the ground states are $\omega_{c}-2\xi\cos \frac{k'\pi}{8}$,
independent of the coupling strength $g$, which contribute to the
straight transmission lines as shown in Fig.~\ref{fig:smalleta-T(a)}.
Moreover, by comparing Fig.~\ref{fig:smalleta-T(a)} with
Fig.~\ref{fig:smalleta-Tin(b)}, we find that these states make no
contributions to the inelastic scattering because
$\langle \psi_{2} | \hat{a}_{N} b^{\dagger}_{k'} |\psi_{0}\rangle \approx 0$ although $\langle \psi_{0} | \hat{b}_{k'} \hat{a}^{\dagger}_{1} | \psi_{0} \rangle \neq 0$.

Notice that in Fig.~\ref{fig:smalleta-T(a)}, a significant
transmission valley appears near every intersection of two
transmission lines, which implies there are two transmission channels
for the photon. The transmission valley is
from the destructive interference between these two channels: one is
provided by the state $b^{\dagger}_{k'} |\psi_{0}\rangle$, while the
other results from the atom coupling with the non-resonant modes of
the SC\@. The physics is the same as that we analyzed in the case
$\eta \ll \xi \ll \omega_{c}$.

\subsection{Numerical results for larger $\eta/\xi$\label{sec:3B}}

As the parameter $\eta$ increases, the coupling between the SC system
and the left and the right channels increases. In particular, when
$\eta=\xi$, the border between the SC system and the outside channels
disappears, and the length of the SC system is artificial. In this
case, the SC system acts as the scattering region, where the
multi-photon processes occur. In the outside channels, we only
consider the single photon process. This approximation can be
identified by numerically checking the convergence of the
transmittance by choosing a longer length of the SC\@.

\begin{figure}[htbp]
  \centering
  \subfloat[The total transmittance $J_{\text{T}}$.]{
    \includegraphics[width=0.45\textwidth]{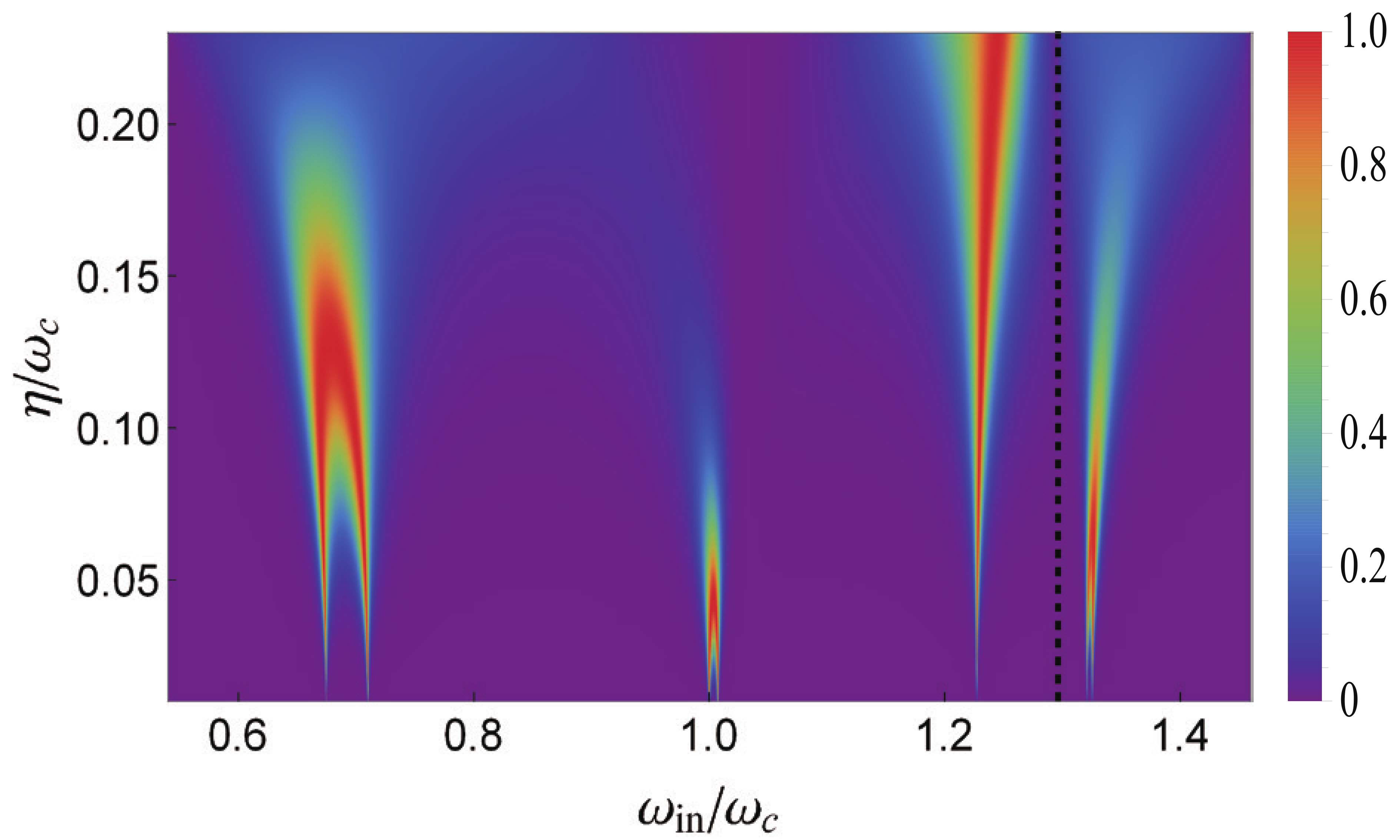}\label{fig:etaT(a)}
  } \\
  \subfloat[The inelastic transmittance $J_{\text{T,in}}$.]{
    \includegraphics[width=0.45\textwidth]{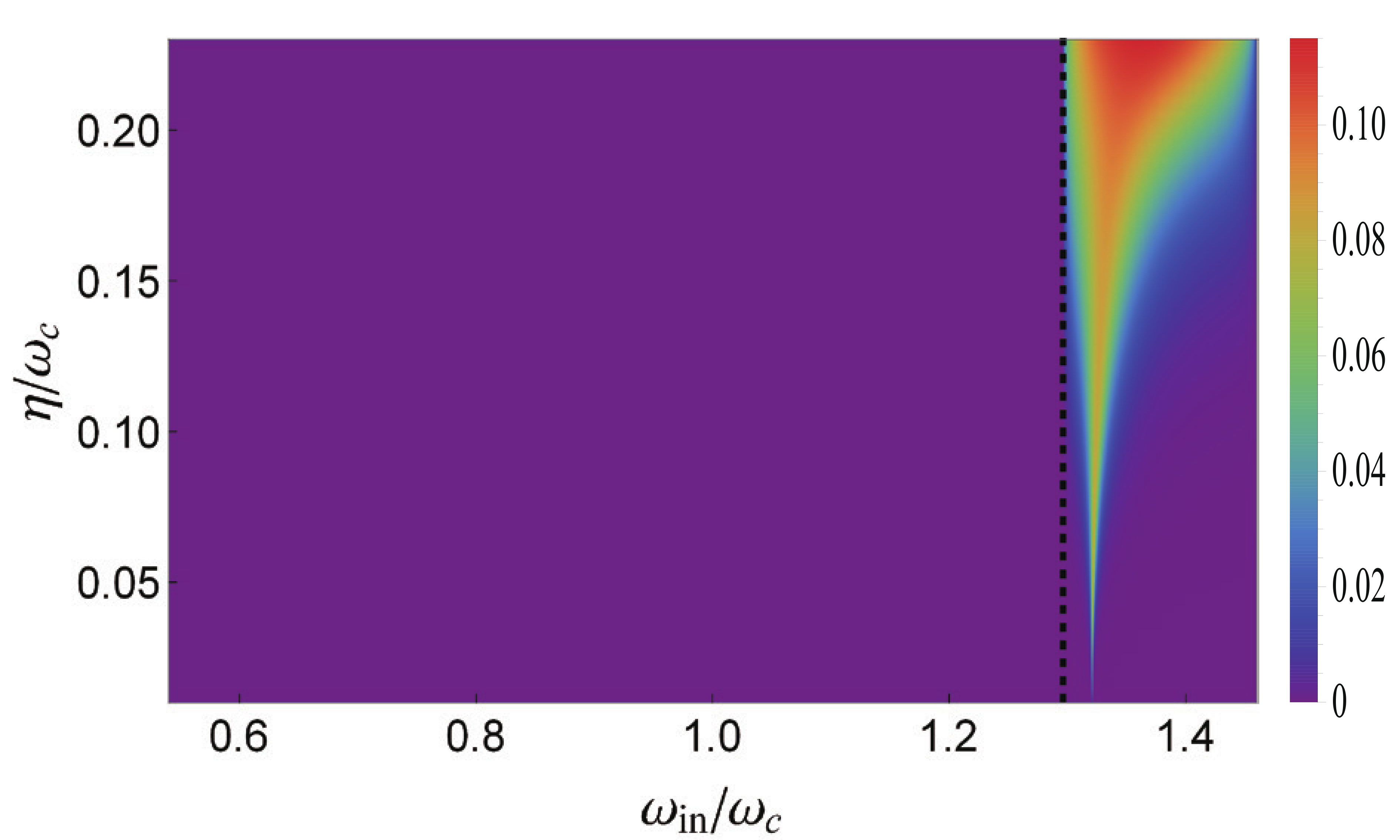}\label{fig:etaTin(b)}
  }
  \caption{(Color online). Transmittance as a function of incident
    photon frequency $\omega_{in}$ and $\eta$. The black dashed line
    marks the frequency given by Eq.~(\ref{eq:9}). Here, we take
    $g/\omega_{c}=0.6$ and choose the length of the SC $N=7$.}
  \label{fig:TeTin-Full-eta}
\end{figure}

In Fig.~\ref{fig:TeTin-Full-eta}, our numerical results for the total
transmittance $J_{\rm{T}}$ and the inelastic transmittance
$J_{\rm{T,in}}$ are plotted in \ref{fig:etaT(a)} and
\ref{fig:etaTin(b)} respectively as functions of $\eta$ and
$\omega_{\text{in}}$ when $g=0.6\omega_{c}$.

As shown in Fig.~\ref{fig:etaT(a)}, the transmission spectrum for
$\eta \ll \xi$ can be explained with the analysis in
Sec.~\ref{sec:3A}. As $\eta$ gets larger, the peak of transmittance
gets wider as expected. When $\eta/\xi$ becomes larger, several peaks
are mixed, and the corresponding eigenstates becomes indistinguishable
from the transmission spectrum. As $\eta/\xi$ becomes sufficient
large, only one transmission peak remains while the other peaks dilute
in the continuous spectrum. This peak implies a resonant state
appearing in our system.

In Fig.~\ref{fig:etaTin(b)}, the inelastic transmittance occurs when
the condition in Eq.~\eqref{eq:8} is satisfied. As $\eta$ is small,
there is an inelastic peak. As $\eta$ increases, the peak develops
into a continuous spectrum.

\begin{figure}[!htbp]
  \centering
  \subfloat[The total transmittance $J_{\text{T}}$.]{
    \includegraphics[width=0.45\textwidth]{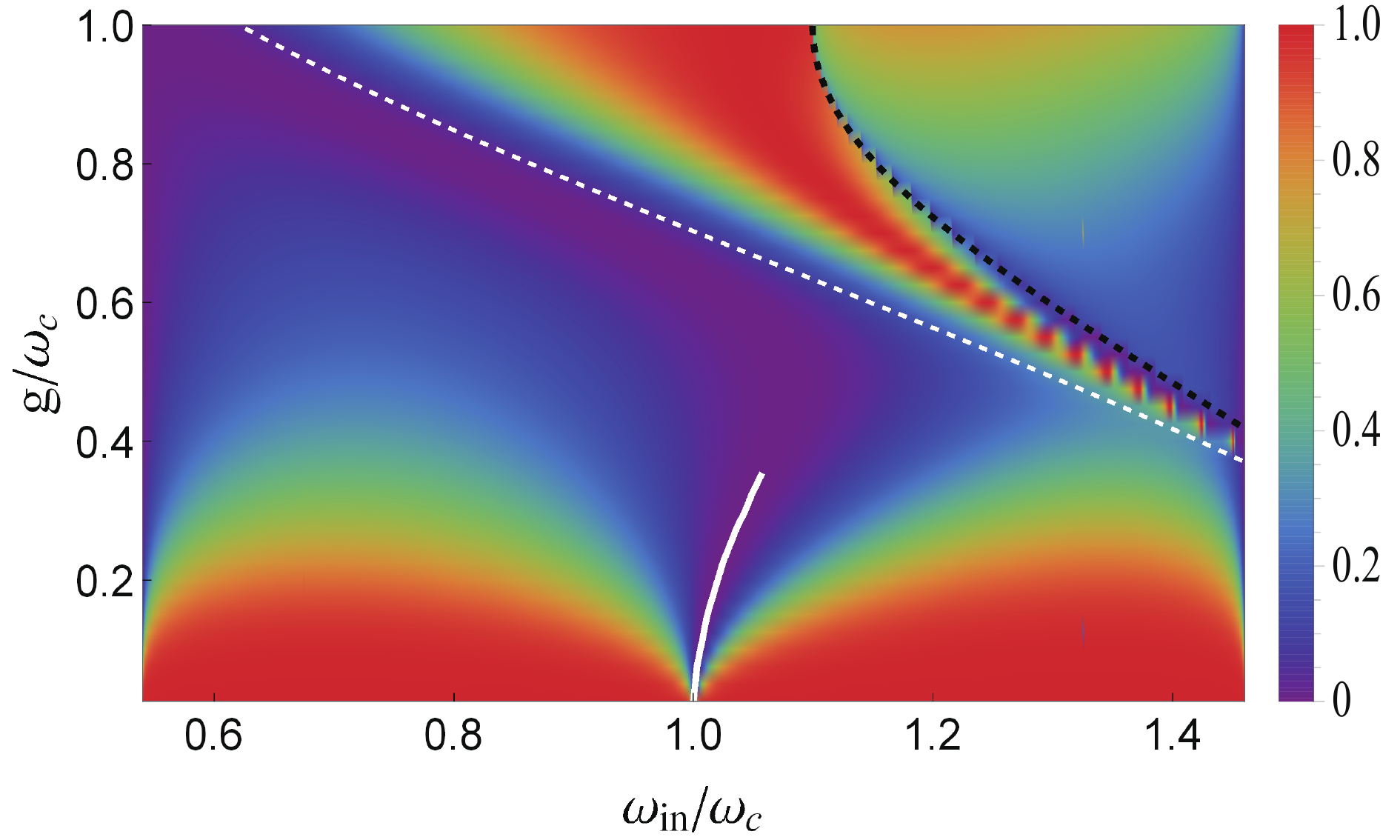}
    \label{fig:eta=xi-T(a)}
  } \\
  \subfloat[The inelastic transmittance $J_{\text{T,in}}$.]{
    \includegraphics[width=0.45\textwidth]{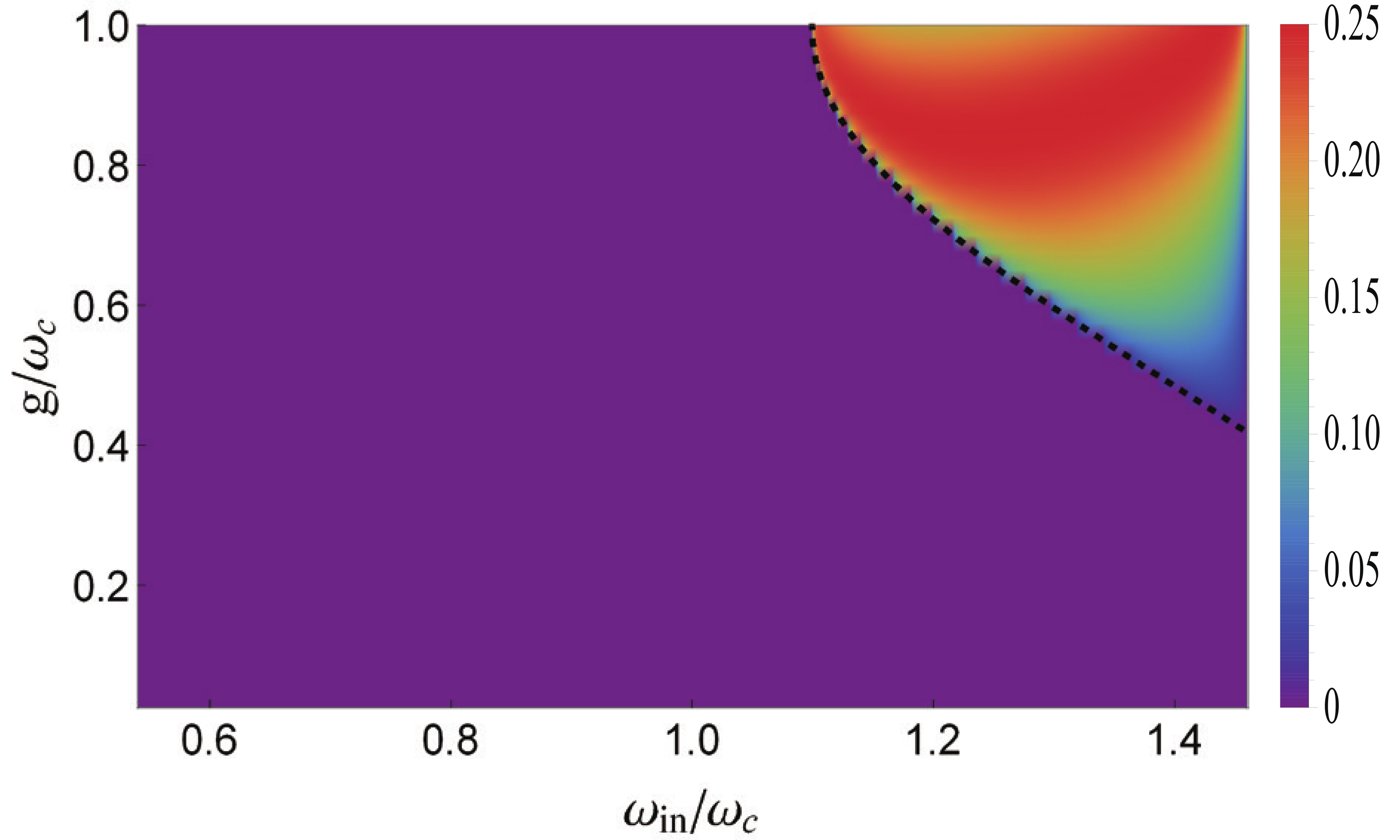}
    \label{fig:eta=xi-Tin(b)}
  }
  \caption{(Color online). Transmittance as a function of incident
    photon frequency $\omega_{in}$ and $g$ for
    $\eta = \xi =0.23\omega_{c}$. The white solid line represents the
    minimum of the elastic transmittance. The white dashed line stands
    for the energy of a bound state, relative to the ground
    state energy $E_{0}$, obtained via the numerical diagnolization.
    And the black dashed line marks the estimated frequency given by
    Eq.~\eqref{eq:8}.}
  \label{fig:TeTin-Full-eta-xi}
\end{figure}

In order to study the properties of transmittance for larger
$\eta / \xi$, we choose $\eta = \xi$ to plot the total transmittance
$J_{\rm{T}}$ and the inelastic transmittance $J_{\rm{T,in}}$ in
Figs.~\ref{fig:eta=xi-T(a)} and \ref{fig:eta=xi-Tin(b)} respectively.

If the RWA is introduced to the model, the minimum of the elastic
transmittance is at $\omega_{min}=\omega_{c}$. However, for
sufficiently large $g/\omega_{c}$, the CRW term will cause the shift
of the minimum frequency which is shown by a white solid line in
Fig.~\ref{fig:eta=xi-T(a)}. This phenomenon has also been observed in
Ref.~\cite{PhysRevLett.113.263604}.

Applying the exact diagonalization in the subspace with the excitation
number $N_{\text{ext}} \geq 3$ of the Hamiltonian~\eqref{eq:2a}, we
obtain a white dashed line in Fig.~\ref{fig:eta=xi-T(a)}, which
represents the energy of a state with odd parity relative to $E_{0}$.
This state is a bound state of the subspace which would be proved in
Appendix~\ref{app:B}. Since it enters into the single-photon
scattering energy regime and couples with single excitation states, it
generates a resonant state~\cite{Atom-PhotonInteractions}, which would
induce the Fano-type resonance~\cite{PhysRev.124.1866} in
Fig.~\ref{fig:eta=xi-T(a)} as mentioned above. Notice that this
resonant state is not a bound state based on Appendix~\ref{app:A},
although its spatial profile of the photon excitations has localized
shape.

In Fig.~\ref{fig:etaT(a)} we demonstrate how a transmission spectrum
at $g/\omega_{c}=0.6$ in Fig.~\ref{fig:smalleta-T(a)} is transformed
into that in Fig.~\ref{fig:eta=xi-T(a)}. The transmission line near
$\omega_{\text{in}}/\omega_{c}\simeq 1.2$ corresponds to the above
resonant quasi-bound state. We find that this quasi-bound state has a
significant component with three excitations, which makes it weakly
coupled with the outside channels, and hence it has a long life time.

Furthermore, the black dashed line in Figs.~\ref{fig:etaTin(b)} and
\ref{fig:eta=xi-Tin(b)} shows the lower energy limit of the incident
photon to observe the inelastic scattering phenomenon. In our case,
the inelastic transmittance never exceeds $25\%$, and the inelastic
reflectance equals to the inelastic transmittance due to the symmetry
of our model.

\section{Conclusion\label{sec:4}}

In this article, we have investigated the single-photon scattering
process via a SC-QED system with the Hamiltonian in Eq.~\eqref{eq:1}
and Eqs.~\eqref{eq:2}. Since the Hamiltonian contains the CRW term in
Eq.~\eqref{eq:5}, it is suitable to study the physics of the
ultrastrong coupling regime. In our study, the parameter $\eta$ varies
in the region $0<\eta/\xi\le 1$. When $\eta/\xi\ll 1$, it describes a
SC-QED system. When $\eta=\xi$, it describes a waveguide QED system.
We find that the condition for the single-photon scattering is
satisfied in our model, which gives us a good opportunity to study the
crossover between these two regimes. In these two regimes, we have
studied how the coupling between the atom and one cavity affects the
transmission. Technically, we present a time independent scattering
theory to describe these single-photon scattering processes, in which
the bound states in the scattering region play an important role. In
the microscopic mechanism to give rise to the inelastic process or to
produce the Fano-type resonance, the bound states or the quasi-bound
states play an essential role.

More precisely, we have predicted the following phenomena for the
transmission spectra. As shown in Fig.~\ref{fig:Cgsmallxi}, the CRW
contribution could be detected even in the weak atom-cavity coupling
regime when the atom is at the node of the resonant of the empty SC
system for $\eta \ll \xi \ll \omega_{c}$. Besides, the CRW induced
inelastic scattering in Figs.~\ref{fig:smalleta-Tin(b)},
\ref{fig:etaTin(b)} and \ref{fig:eta=xi-Tin(b)} will not appear until
Eq.~\eqref{eq:8} is satisfied. By tuning the ratio $\eta/\xi$, we can
take Fig.~\ref{fig:TeTin-Full-eta} as an example to investigate the
single-photon scattering problem in the crossover from a SC-QED
($\eta \ll \xi$) to a waveguide QED ($\eta = \xi$). For the case that
$\eta = \xi$, the blueshift of the elastic transmittance minimum,
which has also been observed in Ref.~\cite{PhysRevLett.113.263604},
can be obtained based on our proposed mechanism. Meanwhile, the
Fano-type resonance~\cite{PhysRev.124.1866} in
Fig.~\ref{fig:eta=xi-T(a)} has been interpreted as the result of a
long-lived quasi-bound state. Further more, the inelastic scattering
phenomena can be obviously observed for sufficiently large
$g/\omega_{c}$.

In summary, we present a unified framework to study the single photon
transmission phenomena induced by the CRW term
$\hat{H}^{\rm{CRW}}_{\rm{int}}$ in our model for any coupling
strengths $\eta/\xi \in (0,1]$ and $g/\omega_{c}\in [0,1]$. Our
results provide theoretical foundations to manipulate the CRW effects
in the corresponding realistic systems. Besides,
although the single-photon scattering condition is satisfied, the
multi-photon processes in the scattering region play a key role in the
effects from the CRW term. We hope that our work will stimulate the
further studies on multi-photon scattering effects induced by the CRW
interaction in many diverse systems.

\begin{acknowledgments}
  This work is supported by NSF of China (Grant Nos. 11475254 and
  11404021) and NKBRSF of China (Grant Nos. 2012CB922104 and
  2014CB921202).
\end{acknowledgments}

\appendix

\section{
Bound states analyzed with BWPT~\label{app:A}}

\subsection{The origin of bound states~\label{app:A1}}

To study the bound state, we have resort to Brillouin-Wigner
perturbation theory~\cite{BWPT2010} instead of the
Rayleigh-Schr$\ddot{o}$dinger perturbation
theory~\cite{Sakurai2011Modern} in that the former will essentially
avoid the possible divergences.

First we divide the Hamiltonian~(\ref{eq:2a}) into two parts,
$\hat{H}_{S}=\hat{H}_{0}+\hat{V}$, where
\begin{equation}
  \hat{H}_{0} = \omega_{c} \sum^{N}_{j=1} \hat{a}^{\dagger}_{j}
  \hat{a}_{j} + \frac{\omega_{a}}{2} \hat{\sigma}_{z} + g
  \hat{\sigma}_{x} (\hat{a}^{\dagger}_{s} + \hat{a}_{s})
  \label{eq:A1}
\end{equation}
is the free Hamiltonian, and
\begin{equation}
  \hat{V} = -\xi \sum^{N}_{j=2} (\hat{a}^{\dagger}_{j}
  \hat{a}_{j-1} + \hat{a}^{\dagger}_{j-1} \hat{a}_{j})
  \label{eq:A2}
\end{equation}
is treated as a perturbation.

In the free Hamiltonian $\hat{H}_{0}$, the $s$-th cavity and the
two-level atom forms the Rabi model, which has been analytically
solved recently in Ref.~\cite{PhysRevLett.107.100401}. Most
eigenstates of $\hat{H}_{0}$ are highly degenerate, and the
non-degenerate eigenstates are given by
\begin{equation}
  |\psi_{m, P}^{(0)}\rangle = |\phi_{m, P}\rangle \otimes_{j=1}^{s-1}
  |0\rangle_{j} \otimes_{j=s+1}^{N} |0\rangle_{j}
  \label{eq:A3}
\end{equation}
for $m\in\{1,2,\cdots\}$, where $|\phi_{m, P}\rangle$ is the $m$-th
eigenstate in the subspace with even parity ($P=1$) or odd parity
($P=-1$) of the Rabi model, and $|0\rangle_{j}$ is the state with $0$
photon in the $j$-th cavity.

Note that the bound states here are the eigenstates of $H_{S}$ where
the photon excitation is localized near the middle of the cavity
array. In this sense the zero order eigenstates
$\{|\psi_{m,P}^{(0)}\rangle\}$ are bound states when $\xi=0$. Since
$\{|\psi_{m,P}^{0}\rangle\}$ are non-degenerate, it is reasonable for
us to expect that when $\xi$ is not very large, the corresponding
eigenstates $\{|\psi_{m,P}\rangle\}$ are still bound states with
parity unchanged.

\subsection{Examination of convergence of numerical results\label{app:A2}}

\begin{figure}[!htbp]
\centering
\includegraphics[width=0.48\textwidth]{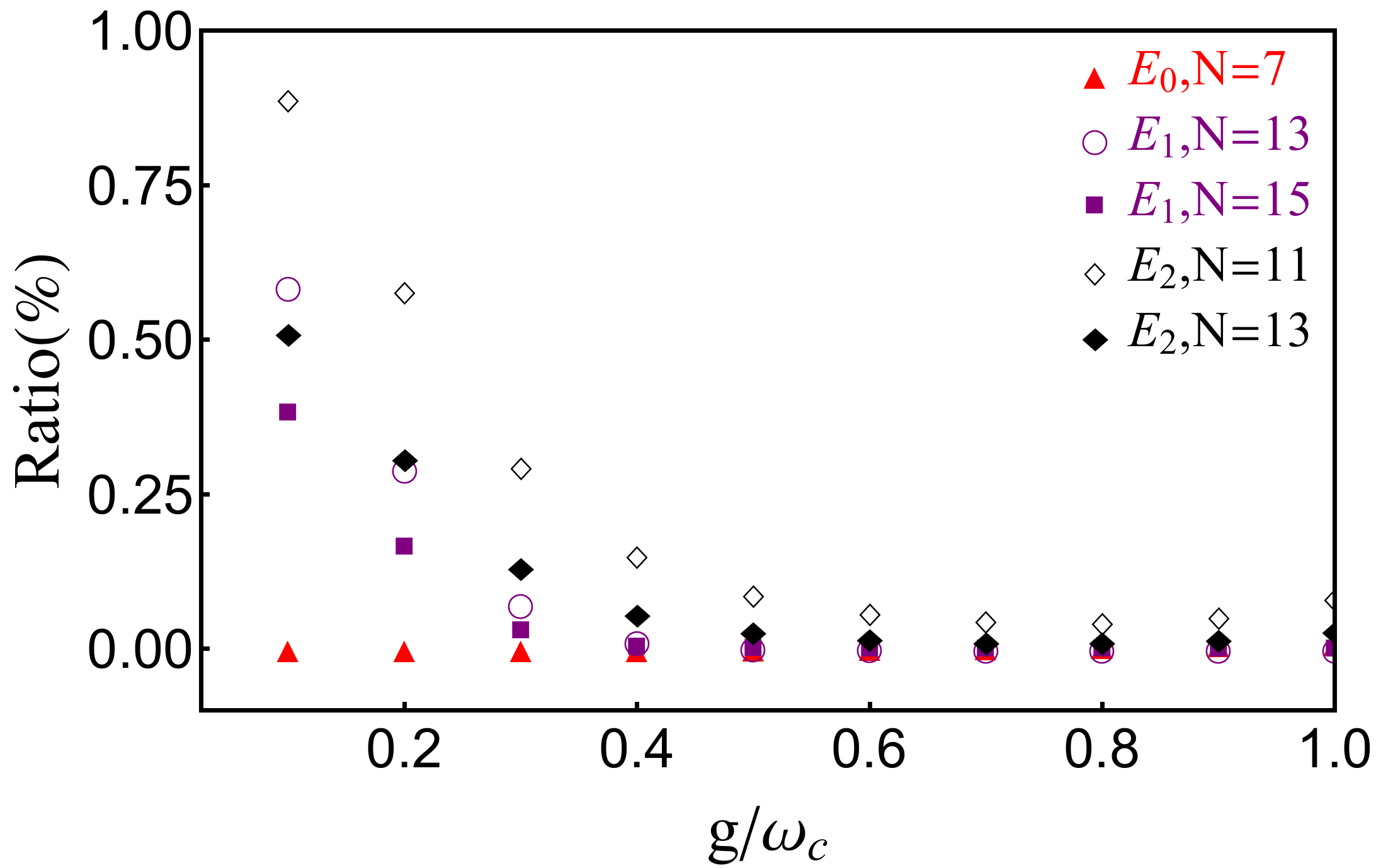}
\caption{(Color online). The ratio of the photon number in the $N$-th
cavity to the total photon excitation number of the bound states as a
function of the coupling strength $g$ for different length $N$ of the
SC. Here we show the results of the ground state with $N=7$ (red
filled triangles), the first excited bound state with $N=13$ (purple
empty circles) and $N=15$ (purple filled squares), the second excited
bound state with $N=9$ (black empty diamonds) and $N=11$ (black filled
diamonds).}
\label{fig:Ratio}
\end{figure}

As we know, when the length $N$ of the SC is long enough, the energy
of a bound state in the SC system should be almost independent of $N$.
Considering the constraint of computational resources, we have to find
a proper $N$ to obtain the highly accurate energy of the bound state,
which implies the necessity of examinating the convergence of our
results. Here, we take the ratio of the photon number in the $N$-th
cavity to the total photon excitation number as reference and show
this ratio as a function of coupling strength $g$ for different length
$N$ in Fig.~\ref{fig:Ratio}.

As shown in Fig.~\ref{fig:Ratio}, the ratio equals to zero for the
ground state, which indicates that $N=7$ is enough to obtain $E_{0}$.
We also find that for higher energy levels, the ratio decreases with
the increase of $g$ and $N$. In order to guarantee the convergence, we
use the BWPT to get $E_{1}$ with $N=15$ and $E_{2}$ with $N=13$ in the
main text Fig.~\ref{fig:BSenergy} to ensure the ratio less than $1\%$.

\section{The bound state of the subspace\label{app:B}}
\begin{figure}[!htbp]
  \centering
  \includegraphics[width=0.45\textwidth]{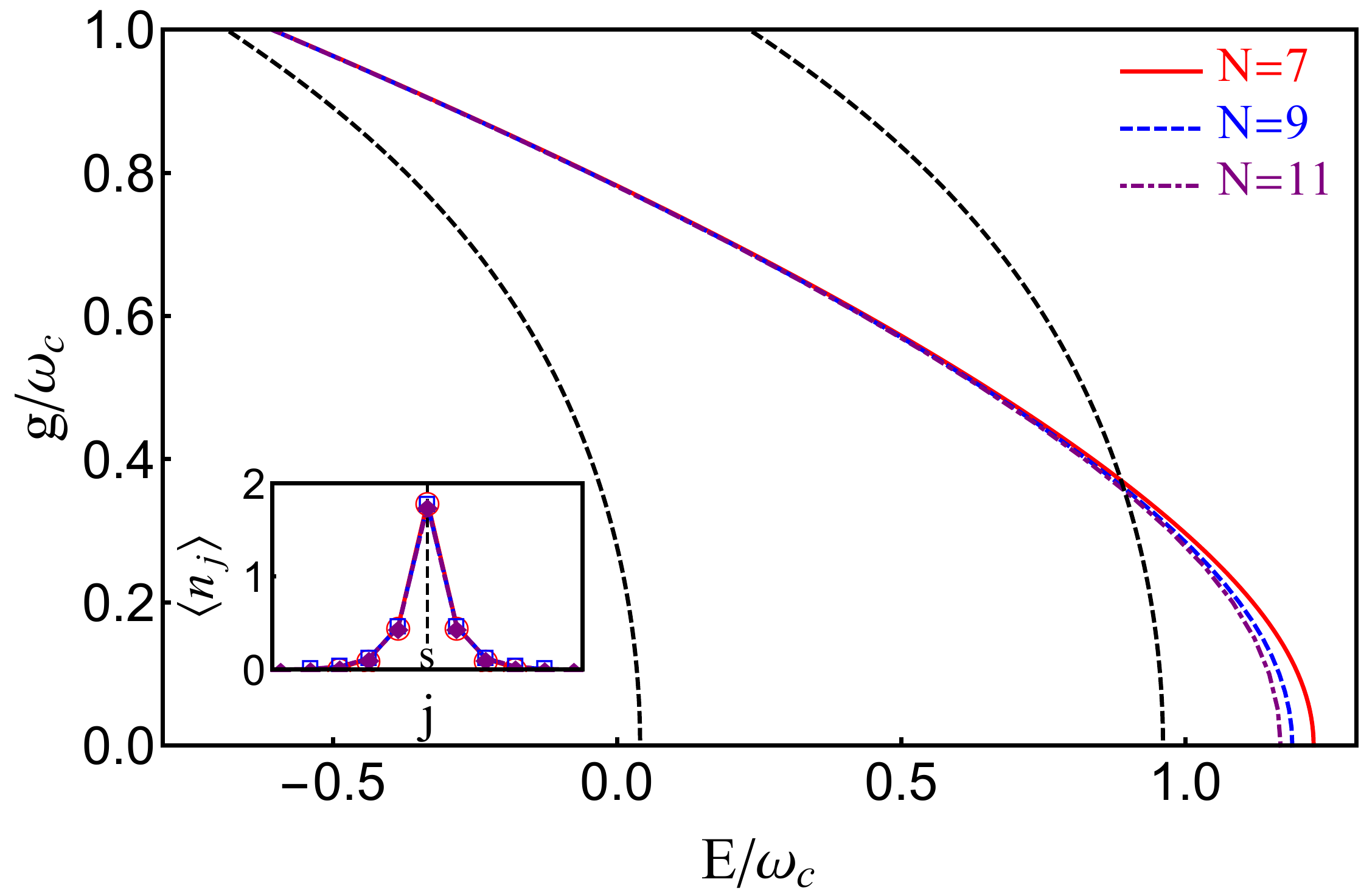}
  \caption{(Color online). The lowest eigenenergy of the subspace with
    the excitation number $N_{\text{ext}} \geq 3$. Dependence with
    coupling strength $g$ and the length $N$ of the SC. The red solid
    line, the blue dashed line and the purple dash-dotted line
    represent respectively results for $N=7$, $N=9$ and $N=11$. The
    single-photon scattering energy regime is the range between two
    black dashed lines. The inset shows the spacial profile of the
    photon excitations in the state, for $g = 0.6\omega_{c}$.}
  \label{fig:BSsub}
\end{figure}

In the subspace with the excitation number $N_{\text{ext}} \geq 3$ of
the Hamiltonian~\eqref{eq:2a}, we can obtain a state with the lowest
energy via numerical diagonalization. The energy of this state as a
function of coupling strength $g$ for different length of the SC are
shown in Fig.~\ref{fig:BSsub}.

As shown in Fig.~\ref{fig:BSsub}, when the coupling strength $g$ is
large enough, the variation of energies with $g$ becomes independent
of the length $N$ of the SC, and its spatial profile of the photon
excitations has localized shape, which implies that it's a bound state
in this regime.

\newpage
\bibliography{reference}

\end{document}